\documentclass{article}
\usepackage{spconf,amsmath,graphicx}

\usepackage{hyperref}
\usepackage{cleveref}

\usepackage{multirow}
\usepackage{multicol}
\usepackage{threeparttable}
\usepackage{setspace}
\usepackage{booktabs}

\title{SA-Paraformer: Non-autoregressive End-to-End Speaker-Attributed ASR}
\name{Yangze Li$^1$, Fan Yu$^2$, Yuhao Liang$^1$, Pengcheng Guo$^1$, Mohan Shi$^3$, Zhihao Du$^2$, Shiliang Zhang$^2$, Lei Xie$^{1*}$\thanks{* Corresponding author.}}
\address{$^1$Audio, Speech and Language Processing Group (ASLP@NPU), Northwestern Polytechnical University \\ $^2$Speech Lab of DAMO Academy, Alibaba Group, $^3$University of Science and Technology of China}
\copyrightnotice{979-8-3503-0689-7/23/\$31.00~\copyright2023 IEEE}

\begin{document}
%
\maketitle
\vspace{-0.6cm}
\begin{abstract}
Joint modeling of multi-speaker ASR and speaker diarization has recently shown promising results in speaker-attributed automatic speech recognition (SA-ASR).
Although being able to obtain state-of-the-art (SOTA) performance, most of the studies are based on an autoregressive (AR) decoder which generates tokens one-by-one and results in a large real-time factor (RTF).
To speed up inference, we introduce a recently proposed non-autoregressive model \textit{Paraformer} as an acoustic model in the SA-ASR model.
Paraformer uses a single-step decoder to enable parallel generation, obtaining comparable performance to the SOTA AR transformer models.
Besides, we propose a speaker-filling strategy to reduce speaker identification errors and adopt an inter-CTC strategy to enhance the encoder's ability in acoustic modeling.
Experiments on the AliMeeting corpus show that our model outperforms the cascaded SA-ASR model by a 6.1\% relative speaker-dependent character error rate (SD-CER) reduction on the test set.
Moreover, our model achieves a comparable SD-CER of 34.8\% with only 1/10 RTF compared with the SOTA joint AR SA-ASR model.
\end{abstract}
\vspace{-0.3cm}
\section{Introduction}
\label{sec:intro}
\vspace{-0.3cm}

Speaker-attributed automatic speech recognition (SA-ASR) is the primary task of multi-speaker speech recognition~\cite{fiscus2006rich,fiscus2007rich,BarkerWVT18,watanabe20b_chime,ryant2020third,mccowan2005ami}.
Besides predicting the corresponding transcriptions of each speaker in overlapping speech~\cite{yu2017recognizing,chen2017progressive,kanda2020serialized,guo2021multi,yu2022mfcca}, SA-ASR also aims to assign speaker labels to the transcriptions, providing users with richer metadata and improved readability.
The multi-party meeting SA-ASR task is considered to be one of the most challenging and valuable tasks in speech applications due to the complex acoustic conditions, such as overlapping speech, an unknown number of speakers, far-field recorded signals, various types of noises, etc~\cite{mccowan2005ami,Yu2022M2MeT,Yu2022Summary}.
As a result, SA-ASR needs to combine multiple related speech processing modules, such as speech separation~\cite{yu2017permutation,hershey2016deep,chen2017deep} to extract speaker representation, speaker diarization~\cite{park2022review,fujita2019end2,horiguchi2020end} to assign speaker label and ASR~\cite{DBLP:journals/corr/VaswaniSPUJGKP17,gulati2020conformer,li2021recent} to transcribe speech.

Substantial research efforts have been dedicated to SA-ASR, which can be mainly classified into two categories: cascaded and joint training approaches.
Most of the cascaded models achieve the goal by combining speech separation, speaker diarization, and speech recognition in a single pipeline~\cite{chen2020continuous,chang2019mimo,zhang2021end,wu2020end}.
Besides, there are also some works based on the serialized output training (SOT) strategy~\cite{kanda2020serialized}, which eliminates an explicit speech separation module and directly predicts multiple outputs from mixed speech.
To obtain speaker-attributed transcriptions, frame-level diarization with SOT (FD-SOT)~\cite{Yu2022ACS} simply aligns the timestamps of the speaker-diarization results and the recognized hypotheses of ASR.
Due to the erroneous timestamps of the alignment strategy, word-level diarization with SOT (WD-SOT)~\cite{Yu2022ACS} approach was proposed to get rid of such alignment dependency by introducing a word-level diarization model.
However, cascaded SA-ASR approaches optimize multiple modules separately which suffers from error propagation and limits their accuracy.

To mitigate such sub-optimality, joint training approaches have been proposed with a promising result~\cite{kanda2021investigation, KandaGWMCZY20,kanda2021comparative}, which integrate multi-speaker ASR and speaker diarization modules into an end-to-end neural solution.
In~\cite{kanda21b_interspeech}, Kanda \textit{et al}. showed a SOTA performance on various multi-talker test sets with an end-to-end SA-ASR model which is built on SOT. 
In detail, they introduce an auxiliary input speaker inventory to produce speaker labels and multi-speaker transcriptions.
However, the autoregressive (AR) architecture of their model recursively generates the next token conditioned on the previously generated tokens, which results in a complex computation and a large real-time factor (RTF) as the sequence length increases.

In contrast, the non-autoregressive (NAR) models~\cite{higuchi2020mask,dong2020cif,gao2022paraformer} aim to perform parallel inference and no longer rely on the left-to-right temporal dependency.
There are two main categories of non-autoregressive ASR models: 
One category is the multi-pass non-self-regression model like mask CTC~\cite{higuchi2020mask}. These models require multiple iterations of the decoder for correction.
As our decoder is very complex, multiple iterations will bring a lot of extra computation.
Another category is the one-pass non-self-regression model represented by the recently-proposed Paraformer~\cite{gao2022paraformer}, which is adopted as our acoustic model because of its superior performance. 
The Paraformer introduces a predictor and glancing language model (GLM) sampler.
The predictor utilizes a continuous integrate-and-fire mechanism to predict the number of tokens accurately and the sampler enhances the decoder's capability to model context information by generating semantic embeddings.
Notably, the Paraformer obtained the best performance on Chinese corpus compared to other NAR models like mask CTC, LASO, TSNAT, etc~\cite{gao2022paraformer}. 
Therefore, we introduce Paraformer in SA-ASR model~\cite{kanda21b_interspeech} as the acoustic model, namely SA-Paraformer.
Following the work in~\cite{kanda21b_interspeech}, we use a SpeakerDecoder which takes in the output of the Paraformer's encoder and predictor as well as the speaker encoder to generate the speaker profile and identifies the speaker of each token.
The decoder of Paraformer then takes in the speaker profile as an auxiliary input to predict the target tokens.
To be able to recognize multi-speaker transcripts simultaneously, we follow the t-SOT~\cite{kanda22_interspeech} strategy.
Besides, since the performance of our NAR SA-Paramformer model is particularly dependent on acoustic representations, we apply the recently proposed inter CTC~\cite{lee2021intermediate} to enhance the representation of the acoustic information, which is attached to an intermediate layer of an acoustic encoder.
Finally, due to the lack of contextual information, our model has more difficulty in extracting the speaker embedding to the corresponding tokens.
To better estimate the speaker identities, we fill redundant speakers' profile distance with random values during training and introduce interference speakers.

Experiments on the AliMeeting corpus show that our proposed SA-Paraformer model outperforms the cascaded SA-ASR (WD-SOT) model by a 6.1\% relative speaker-dependent character error rate (SD-CER) reduction on the test set.
Moreover, with a similar model size, our model achieves a comparable SD-CER with only 1/10 RTF compared to the SOTA autoregressive joint SA-ASR model.

\section{Non-autoregressive Paraformer}
\label{paraformer}
\vspace{-0.1cm}

Given an input sequence $\mathbf{X} = \{\mathbf{x}_1, \dots, \mathbf{x}_T\}$, where $T$ means the number of frames, conventional autoregressive (AR) models produce output tokens $\mathbf{Y} = \{y_1, \dots, y_L\}$ as follows:
\vspace{-0.1cm}
\begin{equation}\label{eq:ar}
P_{\text{ar}}(\mathbf{Y}|\mathbf{X}) = \prod_{l=1}^{L}P(y_l|y_{<l},\mathbf{X}),
\vspace{-0.1cm}
\end{equation}
where L refers to the transcription length. These AR models estimate a target token conditioned on both previously generated tokens and the source input sequence, which makes it hard to compute in parallel and causes a large inference latency.
In contrast, non-autoregressive (NAR) models aim to perform parallel decoding without temporal dependence requirements, which can be formulated as:
\vspace{-0.1cm}
\begin{equation}\label{eq:non-ar}
P_{\text{nar}}(\mathbf{Y}|\mathbf{X}) = \prod_{l=1}^{L}P(y_l|\mathbf{X}).
\vspace{-0.1cm}
\end{equation}

\begin{figure*}[t]
\centering
\resizebox{1.0\textwidth}{!}{\includegraphics{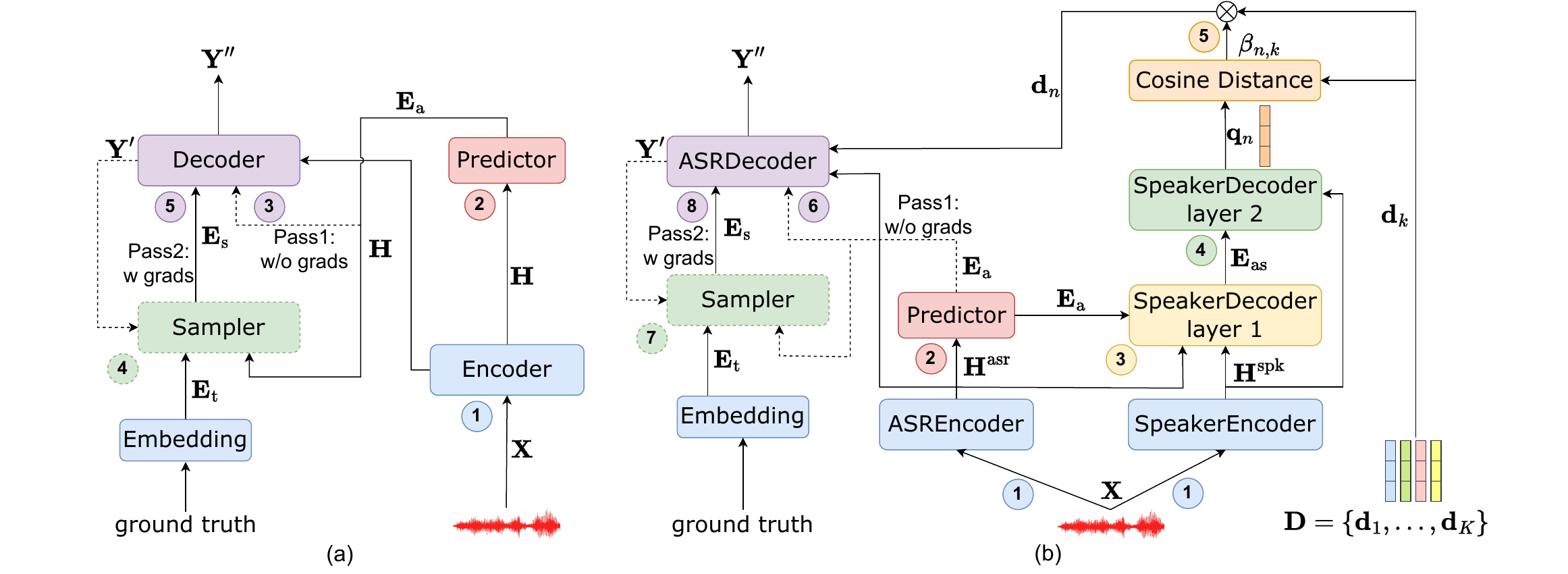}}
\vspace{-0.6cm}
\caption{
 	 (a) Structure of Paraformer. (b) Structure of the proposed SA-Paraformer. 
 	}
  \label{fig1}
\end{figure*}

Recently, an effective NAR model named \textbf{Paraformer}~\cite{gao2022paraformer} was proposed, showing superior performance over other models.
As shown in the left part of Fig.~\ref{fig1}, it mainly integrates two core modules into the base Transformer model, which are the predictor and sampler.

\textbf{Predictor} is used to extract the acoustic embedding $E_{\text{a}}$ corresponding to each token by introducing the mechanism of Continuous Integrate-and-Fire (CIF)~\cite{dong2020cif}.
At each encoder step, the predictor first predicts a weight $\alpha$ for each frame to scale acoustic information and then accumulates $\alpha$  to integrate hidden representations $E_s$ until the accumulated weight reaches a given threshold $\beta$, which indicates that an acoustic boundary has been reached.
The weight $\alpha$ is also accumulated to estimate the output token number, which provides a soft alignment between acoustic frames and target labels.
Moreover, a mean absolute error (MAE) loss is added to improve the accuracy of sequence length prediction as follows:
\vspace{-0.1cm}
\begin{equation}\label{eq:quantity}
\mathcal L_{\text{mae}} = \vert \mathcal{N} - \sum_{t=1}^{T} \alpha_{t} \vert,
\vspace{-0.1cm}
\end{equation}
where $\mathcal{N}$ is the ground-truth length of the target sequence. 

\textbf{Sampler} regenerates a semantic embedding $\mathbf{E}_{\text{s}}$  by randomly replacing acoustic embedding $\mathbf{E}_{\text{a}}$ with char token embedding $\mathbf{E}_{\text{t}}$. 
The number of replacements is determined by the character error number between ground truth transcripts $\mathbf{Y}_{\text{tr}}$ and the first pass hypotheses $\mathbf{Y}^{\prime}$.
\vspace{-0.1cm}
\begin{equation}
    \mathbf{E}_{\text{s}}=\text{Sampler}\left(\mathbf{E}_{\text{a}}, \mathbf{E}_{\text{t}}, \left \lceil \lambda d\left( \mathbf{Y}_{\text{tr}}, \mathbf{Y}^{\prime}\right) \right \rceil \right),
    \vspace{-0.1cm}
    \label{sampler}
\end{equation}
where sampling factor $\lambda$ is used to control the sample ratio and $d\left(\mathbf{Y}_{\text{tr}}, \mathbf{Y}^{\prime}\right)$ is the function of character error number calculation.

\section{Proposed Method}
\label{sec:pagestyle}

\subsection{Model architecture}
Our proposed SA-Paraformer approach aims to obtain the speaker-attributed transcriptions, including transcription $\mathbf{Y}$ and speaker identity $\mathbf{S}$, given feature sequence $\mathbf{X}$ and speaker inventory $\mathbf{D}$ as input.
The speaker inventory $\mathbf{D} = \{\mathbf{d}_1,\dots,\mathbf{d}_K\}$ contains $K$ speaker profile vectors (e.g., d-vector~\cite{variani2014deep}).
The SA-Paraformer model is composed of six modules: ASREncoder, ASRDecoder, SpeakerEncoder, SpeakerDecoder, Predictor, and Sampler, as depicted in Fig.~\ref{fig1}(b).
Firstly, the ASREncoder and SpeakerEncoder are responsible for transforming the input sequence $\mathbf{X}$ into ASR hidden representation $\mathbf{H}^{\text {asr}}$ (Eq. (\ref{eq:asr_enc})) and speaker hidden representation $\mathbf{H}^{\text { spk }} $ (Eq. (\ref{eq:spk_enc})), respectively.
Then, the predictor module takes the ASR hidden representation $\mathbf{H}^{\text {asr}}$ as input and extracts acoustic embedding $\mathbf{E}_{\text{a}}$ corresponding to each token (Eq. (\ref{eq:prdic})).
\begin{align}
    & \mathbf{H}^{\text { asr }} = \text { ASREncoder }(\mathbf{X}) \label{eq:asr_enc}, \\
    & \mathbf{H}^{\text { spk }} = \text { SpeakerEncoder }(\mathbf{X}) \label{eq:spk_enc}, \\
    & \mathbf{E}_\text{a} = \text { Predictor }(\mathbf{H}^{\text { asr }}) \label{eq:prdic}.
\end{align}
SpeakerDecoder is designed to generate a speaker profile $d_n$ for each token and it consists of two Transformer layers and a cosine-distance-based attention mechanism. The first Transformer layer can be described as:
\begin{align}
    & \mathbf{E}^{\prime}_{\text{as}} = \mathbf{E}_{\text{a}} + \mathrm{MHA}^{\text { spk-src }}\left(\mathbf{E}_a, \mathbf{H}^{\text { asr }}, \mathbf{H}^{\text { spk }}\right),  \label{eq:spk_dec_mha} \\
    & \mathbf{E}_{\text{as}} = \mathbf{E}^{\prime}_{\text{as}} + \mathrm{FF}^{spk}\left(\mathbf{E}^{\prime}_{\text{as}}\right) \label{eq:spk_dev_ffn}.
\end{align}
Here, to obtain the speaker embedding of each token $\mathbf{E}^{\prime}_{as}$, we use $\mathbf{E}_a$, $\mathbf{H}^{\text {asr}}$ and $\mathbf{H}^{\text {spk}}$ as the query, key, and value of the multi-head attention (MHA) (Eq. (\ref{eq:spk_dec_mha})).
Then, feed-forward (FF) layer receives $\mathbf{E}^{\prime}_{\text{as}}$ to obtain $\mathbf{E}_{\text{as}}$ (Eq. (\ref{eq:spk_dev_ffn})).
In the second Transformer layer, only the speaker representation $\mathbf{H}^{\text{spk}}$ is used as input, as it already contains sufficient acoustic information for each token (Eq. (\ref{eq:spk_dec_layer2})).
To obtain the speaker query $\mathbf{q}_n$, a linear layer is applied to adapt the dimension to the speaker inventory $\mathbf{D}$, using a learnable weight $\mathbf{W}_{\text{spk}}$ (Eq. (\ref{eq:spk_dec_ln})). The computation of $q_n$ is:
\vspace{-0.1cm}
\begin{align}
    & \mathbf{E}_{\text{spk}} = \text{TransformerLayer}\left(\mathbf{E}_{as}, \mathbf{H}^{\text{spk}}\right) \label{eq:spk_dec_layer2},  \\
    & \mathbf{q}_{n}= \mathbf{W}_{\text{spk}} \cdot \mathbf{E}_{\text{spk}}  \label{eq:spk_dec_ln}.
    \vspace{-0.1cm}
\end{align}
Then, a cosine-distance-based attention weight $\mathbf{b}_{n, k}$ is calculated between each profile $\mathbf{d}_{k}$ in the speaker inventory $\mathbf{D}$ and the obtained speaker query $\mathbf{q}_n$. The attention weight $\beta_{n, k}$ is derived from $\mathbf{b}_{n, k}$ through a SoftMax function. Finally, the attention-weighted speaker profile $\mathbf{d}_n$ is obtained by the dot product of $\beta_{n, k}$ and $\mathbf{D}$:
\vspace{-0.1cm}
\begin{align}
    & \mathbf{b}_{n, k}=\frac{\mathbf{q}_{n} \cdot \mathbf{d}_{k}}{\left|\mathbf{q}_{n}\right|\left|\mathbf{d}_{k}\right|}, \\
    & \beta_{n, k}=\frac{\exp \left(\cos \left(\mathbf{b}_{n,k}, \mathbf{d}_{k}\right)\right)}{\sum_{j}^{K} \exp \left(\cos \left(\mathbf{b}_{n,k}, \mathbf{d}_{j}\right)\right)}, \label{eq:beta}\\
    & {\mathbf{d}}_{n}=\sum_{k=1}^{K} \beta_{n, k} \mathbf{d}_{k}. 
    \vspace{-0.1cm}
\end{align}
ASRDecoder inputs the acoustic embedding $\mathbf{E}_{\text{a}}$, ASR hidden representation $\mathbf{H}^{asr}$ and weighted profile ${\mathbf{d}}_{n}$ to produce the first pass hypotheses $\mathbf{Y}^{\prime}$ without backward gradients. 
The architecture of the ASRDecoder is similar to a Transformer decoder, except that $\mathbf{d}_n$ is added to the feed-forward module at the first layer (Eq. (\ref{asr_decoders_ffn})).
\vspace{-0.1cm}
\begin{align}
    & \mathbf{E}^{\prime}_{c,1}=\mathbf{E}_a+\text{MHA}(\mathbf{E}_a, \mathbf{H}^{\text{asr}}, \mathbf{H}^{\text{asr}}) \label{asr_decoder1}, \\
    & \mathbf{E}_{c,1}=\mathbf{E}^{\prime}_{c,1}+\mathrm{FF}\left(\mathbf{E}^{\prime}_{c,1}+\mathbf{W}_{\text{spk}} \cdot \mathbf{d}_n\right)  \label{asr_decoders_ffn}, \\
    & \mathbf{E}_{c,l}=\text{TransformerLayer}_{l}\left(\mathbf{E}_{c,l-1}, \mathbf{H}^{\text{asr}}\right) \label{asr_decoders}, \\
    & \mathbf{Y}^{\prime}=\text{SoftMax}\left(\mathbf{W}_{\text{out}} \cdot \mathbf{E}_{c,L} + \mathbf{b}_{\text{out}}\right) \label{asr_decoder_out},
    \vspace{-0.1cm}
\end{align}
where $l \in \left(1, L\right)$ denotes the $l$-th layer of the ASRDecoder, $\mathbf{W}_{\text{out}}$ and $\mathbf{b}_{\text{out}}$ are learnable weight and bias parameter, $\text{Softmax}(\cdot)$ is the column-wise softmax function.
Then we adopt sampler, described in Section \ref{paraformer}, to regenerate semantic embedding $\mathbf{E}_{\text{s}}$ (Eq. (\ref{sampler})).
Finally, ASRDecoder takes in semantic embedding $\mathbf{E}_\text{s}$ as well as ASR hidden representations $\mathbf{H}^{\text{asr}}$ to generate the second pass hypotheses $\mathbf{Y}^{\prime \prime}$ with backward gradients.

\subsection{Speaker filling}
Joint SA-ASR models, including our proposed SA-Paraformer model, are independent of the number of speakers in overlapping speech segments.
But in practice, the unknown number of speakers is one of the challenges of multi-party meeting transcription, and the performance of SA-ASR models is easily affected by the number of speakers~\cite{KandaGWMCZY20}.
The interfering speaker (i-speaker) approach was used~\cite{KandaGWMCZY20} to improve the robustness of the model for different speaker numbers, in which several interfering speaker profiles are added into  speaker inventory $D$ of each utterance.
However, our model uses $\mathbf{E}_\text{a}$ to predict speaker profiles that lack contextual information, making speaker identification more difficult.
Therefore, we propose a filling speaker (f-speaker) strategy to further improve the generalization of our model by applying random disturbance.
Specifically, we first expand the speaker profile number to the maximum speakers' number of samples in a batch. Then filling the cosine distance $b_{n, k}$ of the redundant speakers with $-0.5$ to $0.5$ instead of negative infinity.

\subsection{Training strategy}
\subsubsection{Loss function}
\label{training}
During training, all the network parameters are optimized by four loss functions, which are MAE, cross-entropy (CE), CTC, and speaker loss. Especially, the speaker loss is added to help the model identify speakers. Thus, the final loss becomes:
\vspace{-0.1cm}
\begin{equation}
    \mathcal{L}=\mathcal{L}_{\text{M A E}}+\lambda_1 \cdot \mathcal{L}_{\text{CTC}}+\left(1-\lambda_1\right) \cdot \mathcal{L}_{\text{CE}}+\mathcal{L}_{\text{spk}},
    \vspace{-0.1cm}
    \label{loss}
\end{equation}
where $\lambda_1$ is an interpolation factor which is set to $0.3$ in this study. Speaker loss function $\mathcal{L}_{spk}$ is defined as:
\vspace{-0.1cm}
\begin{equation}
    \mathcal{L}_{\text{spk}}=\sum_{ n }^{}-\log\frac{e^{b_{n,i}}}{\sum_{k}^{}e^{b_{n,k}}},
    \vspace{-0.1cm}
\end{equation}
where $b_{n,i}$ means the cosine-distance-based attention weight of the $i$-th speaker. To further enhance the representation of the acoustic information, we introduce the inter CTC loss~\cite{lee2021intermediate} into the intermediate layer of the acoustic encoder, expanding Eq.\eqref{loss} as:
\vspace{-0.1cm}
\begin{equation}
    \begin{split}
        \mathcal{L}^{\prime}=\mathcal{L}_{\text{M A E}}+\lambda_1 \cdot \mathcal{L}_{\text{CTC}}+\lambda_2 \cdot \mathcal{L}_{\text{interCTC}} \\ 
                    +\left(1-\lambda_1-\lambda_2\right) \cdot \mathcal{L}_{\text{CE}}+\mathcal{L}_{\text{spk}},
    \end{split}
    \vspace{-0.1cm}
    \label{L'}
\end{equation}
where $\lambda_2$ is also an interpolation factor which is set to $0.3$ in this study. 

\subsubsection{Token-level serialized output training (t-SOT)}
Serialized output training (SOT)~\cite{kanda2020serialized} strategy is recently proposed to generate transcriptions for multiple speakers in an effective and simple way.
In the SOT strategy, transcriptions of different speakers are serialized by their start time which does not match the assumption of temporal monotonicity.
Therefore, in this paper, we introduce token-level serialized output training (t-SOT)~\cite{kanda22_interspeech} strategy to enable the multi-speaker recognition ability, which is more friendly to the temporal monotonicity of CTC and Paraformer predictor in our SA-Paraformer model.
The t-SOT strategy generates tokens of multiple speakers in chronological order based on token end times.
However, a special token separator $\left\langle \text{cc} \right\rangle$ is inserted between tokens for channel changes, which does not contain actual acoustic information.
So our proposed SA-Paraformer model is difficult to estimate the acoustic boundary of separator $\left\langle \text{cc} \right\rangle$.
We experimentally investigate training our model with/without the separator $\left\langle \text{cc} \right\rangle$ in section~\ref{separator}.

\section{Experiments}
\label{Experiments}

\subsection{Datasets}
We use AliMeeting corpus~\cite{Yu2022M2MeT,Yu2022Summary}, a challenging Mandarin meeting dataset with multi-talker conversations, to evaluate our proposed SA-Parafomer approach.
The AliMeeting corpus contains 104.75 hours of the training set(Train), 4 hours of the evaluation set(Eval), and 10 hours of the testing set(Test).
The meeting sessions of each set consist of a 15 to 30-minute discussion by a group of participants.
The average speech overlap ratio of the Train set is 42.27\%, which contains a number of multi-talker discussions.
In this work, we use the first channel of the far-field data.

\subsection{Experimental setup}
\label{setting}
In all experiments, we use the 80-dimensional Mel-filterbank feature with an 8 ms window shift and a 32 ms frame length.
For the speaker profile, we employ a 256-dim Res2Net-based d-vector extractor~\cite{variani2014deep} trained on the VoxCeleb corpus~\cite{nagrani2017voxceleb}. As we transcribe for conference audio, we got speaker profiles by clustering.

Our ASR module is comprised of 12 layers of conformer encoder~\cite{gulati2020conformer} and 6 layers of transformer decoder. 
SpeakerEncoder is the same as the d-vector extractor.
The dimension of attention and feed-forward layers are set to 256 and 2048, respectively.
For the training process, we first trained a speaker-agnostic multi-speaker Paraformer model with the Adam optimizer. 
Then we use the well-trained Paraformer model and d-vector extractor as initialization.
All the experiments are conducted on NVIDIA RTX 3090 GPUs and measured by speaker-dependent character error rate (SD-CER)~\cite{Yu2022ACS}, which is calculated by comparing the ASR hypothesis and the reference transcription of the corresponding speaker.
We measure RTF using single-threaded bar-by-bar decoding on an Intel(R) Xeon(R) CPU E5-2678 v3 @ 2.50GHz and an NVIDIA GeForce RTX 2080 Ti, respectively.

\begin{table}[!htb]
\centering
\caption{SD-CER of various SA-ASR approaches on Eval and Test sets. Real-time factor (RTF) is computed as the ratio of the total inference time to the total duration of the Test set.}
\setlength{\tabcolsep}{3.5pt}
\begin{threeparttable}
\begin{tabular}{lcccc}
\toprule
\hline
\multicolumn{1}{c}{\multirow{2}{*}{Approach}} & \multicolumn{2}{c}{SD-CER (\%)}                          & \multicolumn{2}{c}{RTF}                           \\ \cmidrule(r){2-3} \cmidrule(r){4-5}   
\multicolumn{1}{c}{}                          & \multicolumn{1}{c}{Eval} & \multicolumn{1}{c}{Test} & \multicolumn{1}{c}{CPU} & \multicolumn{1}{c}{GPU} \\ \hline
\textit{Cascaded SA-ASR}                    & \multicolumn{1}{c}{}    & \multicolumn{1}{c}{}    & \multicolumn{1}{c}{} & \multicolumn{1}{c}{}  \\
\quad FD-SOT~\cite{Yu2022ACS}              & 41.0      & 41.2  & - & - \\
\quad WD-SOT~\cite{Yu2022ACS}              & 36.0      & 37.1  & - & - \\ \hline
\textit{Joint SA-ASR}                    & \multicolumn{1}{c}{}    & \multicolumn{1}{c}{}    & \multicolumn{1}{c}{} & \multicolumn{1}{c}{}  \\
\quad AR SA-ASR~\cite{shi2023casa}              & \textbf{31.8}      & \textbf{34.7}  & 0.967 & 0.315 \\
\quad SA-Paraformer       & 36.2      & 38.6  &  &    \\
\quad \quad  + interCTC    & 34.5      & 36.9 &  &     \\
\quad \quad  \quad + f-speaker     & 33.3      & 35.7 & \textbf{0.168} & \textbf{0.032}    \\
\quad \quad \quad + i-speaker & 33.1      & 35.6  &  &    \\
\quad \quad \quad + f\&i-speaker & 32.5  & 34.8  &  &   \\
\hline
\bottomrule
\end{tabular}
\end{threeparttable}
\label{tab:general_results}
\vspace{-0.2cm}
\end{table}

\subsection{Comparison of different SA-ASR approaches}
As shown in Table~\ref{tab:general_results}, our proposed SA-Paraformer approach outperforms the modular SA-ASR approaches, especially for the FD-SOT approach, leading to 11.7\% (41.0\% $\to$ 36.2\%) and 6.3\% (41.2\% $\to$ 38.6\%) relative SD-CER reduction on Eval and Test sets, respectively. 
When incorporating with inter-CTC loss, we obtain further improvement, decreasing the SD-CER from 36.2\%/38.6\% to 34.5\%/36.9\% on Eval and Test sets, respectively.
The performance of SA-ASR models is easily affected by the number of speakers. 
Speaker filling (f-speaker) and interfering speakers (i-speaker) strategy can improve the robustness of the model with different speaker number setups.
According to the results, our proposed f-speaker strategy achieves 3.4\% (34.5\% $\to$ 33.3\%) and 3.2\% (36.9\% $\to$ 35.7\%) relative SD-CER reduction on Eval and Test sets.
And i-speaker strategy obtains similar improvements, decreasing the SD-CER from 34.5\%/36.9\% to 33.1\%/35.6\% on Eval and Test sets, respectively.
Finally, combining the f-speaker and i-speaker strategy, our SA-Paraformer approach achieves a comparable SD-CER of 32.5\% and 34.8\% on Eval and Test sets with only 1/10 latency compared with a state-of-the-art autoregressive (AR) joint SA-ASR model which was also initialized based on a pre-trained ASR model.

\vspace{-0.3cm}
\subsection{Impact of the sampling factor}
As shown in Table~\ref{tab:sampler}, we evaluate the sampling factor $\lambda$ in the sampler, described in Section \ref{paraformer}.
Here, $\lambda = 0.0$ means the model training without a sampler mechanism.
When increasing $\lambda$ from 0.7 to 1.1, we observe that the SD-CER is improved from 37.1\% to 36.2\% on the Eval set and 39.2\% to 38.6\% on the Test set, due to the better semantic information provided by the ground truth transcripts during training.
However, when the sampling factor $\lambda$ is too large, it will lead to a mismatch between training and inference, where we decode twice with ground truth transcripts for training and decode once without transcripts for inference~\cite{gao2022paraformer}.
\vspace{-0.2cm}
\begin{table}[!htb]
\centering
\caption{Results of SA-Paraformer model with different sampling factor $\lambda$ on Eval and Test sets (SD-CER \%).}
\begin{tabular}{cccccccccc}
\toprule
\hline
$\lambda$   & 0.0     & 0.7   & 0.9   & 1.1   & 1.3   & 1.5    \\ \hline
Eval        & 38.8    & 37.1  & 36.5  & \textbf{36.2}  & 36.4  & 36.9   \\ 
Test        & 41.1    & 39.2  & 38.8  & \textbf{38.6}  & 38.7  & 39.0   \\
\hline
\bottomrule
\end{tabular}
\label{tab:sampler}
\vspace{-0.1cm}
\end{table}

\vspace{-0.2cm}
\subsection{Comparison of training with/without separator $\left\langle \text{cc} \right\rangle$}
\label{separator}
\begin{figure}[!ht]
\centering
\includegraphics[scale=0.165]{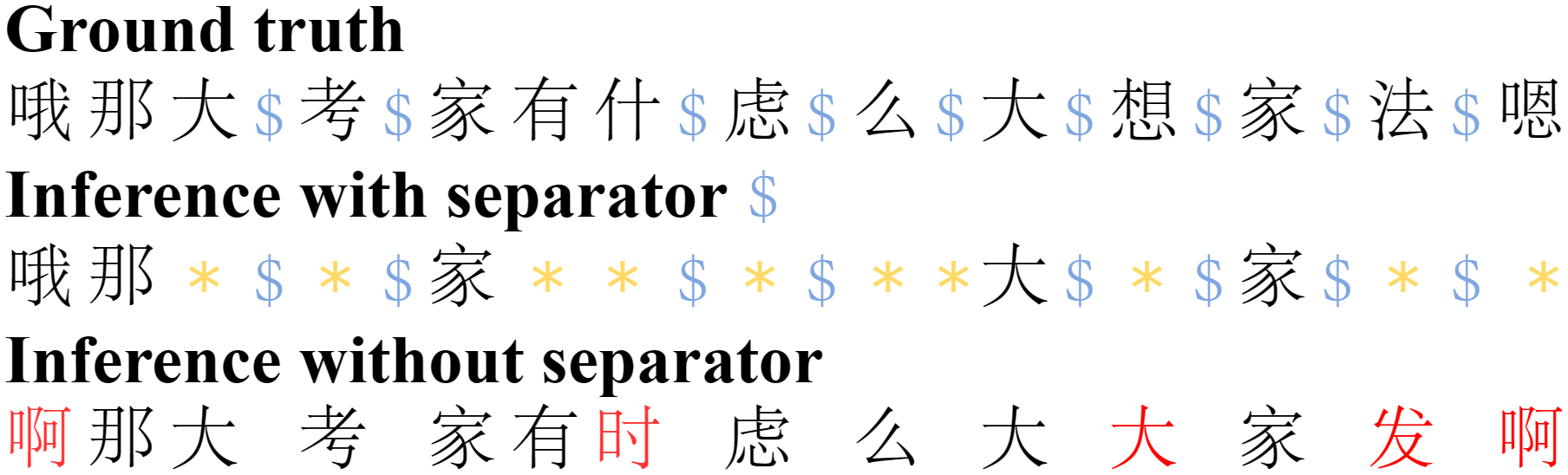}
\caption{
 	 Decoding example for R8002-MS802-0004385-0004672 in the AliMeeting Test set.
            Red indicates substitution error and yellow indicates deletion error.
 	}
  \label{fig2}
\end{figure}

Token separator $\left\langle \text{cc} \right\rangle$ is inserted between tokens for speaker changes, which does not contain actual acoustic information.
So, the predictor of our SA-Paraformer model, based on acoustic boundary estimation, is difficult to deal with the separator $\left\langle \text{cc} \right\rangle$.
As shown in Table~\ref{tab:seperator}, we compared the CER detailed results of the SA-Paraformer model training with (A1 model) and without (A2 model) the separator $\left\langle \text{cc} \right\rangle$. 
Compared with the A1 model, the A2 model brings 10.4\% (43.1\% $\to$ 38.6\%) relative CER reduction on the Test set, due to the decreasing of deletion (Del) errors (26.5\% $\to$ 6.5\%).
As can be seen from Fig 2, the upper part shows the ground truth sequence and the middle part shows the A1 model inference sequence.
After analyzing the decoding results, we find that most of the deletion errors of the A1 model are the normal token.
For the utterance of multi-speaker discussion, the A1 model outputs separator $\left\langle \text{cc} \right\rangle$ and ignores normal token.
We remove the separator during training to improve the prediction accuracy of the normal token boundary for recovering the deletion errors. 
As shown in the lower part of Fig.~\ref{fig2}, the deletion error of the normal token is successfully corrected.
\vspace{-0.3cm}
\begin{table}[!ht]
\centering
\caption{Results of training with/without separator on Test set (CER \%).}
\begin{tabular}{cccccc}
\toprule
\hline
Model & With separator & Ins & Del  & Sub  & CER  \\ \hline
 A1 & Yes           & 3.4 & 26.4 & 13.3 & 43.1 \\
A2 &No            & 4.2 & 6.8  & 27.6 & 38.6 \\ \hline
\bottomrule
\end{tabular}
\label{tab:seperator}
\vspace{-0.1cm}
\end{table}

\vspace{-0.3cm}
\section{Conclusion}
\label{Conclusion}
In this paper, we propose an SA-Paraformer model based on the multi-speaker t-SOT framework to transcribe speech and identify speakers simultaneously, which introduces the Paraformer non-autoregressive model as an acoustic model for parallel decoding.
Besides, we also propose a simple strategy speaker filling to improve the robustness of the model with different
speaker number.
Finally, we introduce the interfering speakers and inter-CTC strategy to obtain further improvement.
Evaluated on the AliMeeting corpus, our proposed SA-Paraformer model achieves 6.1\% relative SD-CER improvement compared with the cascaded SA-ASR model on the test set. 
Moreover, the SA-Paraformer model achieves a comparable SD-CER of 34.8\% to the state-of-the-art autoregressive joint SA-ASR model with 30 times speedup on GPU.
In the future, we would like to integrate a multi-channel model into our proposed SA-Paraformer model for real-world applications.

\bibliographystyle{IEEEtran}
\bibliography{mybib}

\end{document}